\documentstyle[12pt]{article}
\def\ba{\begin{array}}
\def\ea{\end{array}}
\def\be{\begin{equation}\begin{array}{l}}
\def\ee{\end{array}\end{equation}}
\def\bea{\begin{equation}\begin{array}{l}}
\def\eea{\end{array}\end{equation}}
\def\f{\frac}
\def\om{\omega}
\def\omm{\omega^a_b}
\def\we{\wedge}

\def\va{\varepsilon}
\def\omb{\bar{\omega}}
\def\la{\lambda}

 \title{ {\bf On Torsion and Nieh-Yan Form} }

\author{Han-Ying Guo, Ke Wu and Wei Zhang \thanks{zhwei@itp.ac.cn} \\
 Institute of Theoretical Physics , Academia Sinica,\\
 P.O.Box 2735, Beijing, 100080, China  }
\date{ }
\begin{document}
\maketitle
\begin{abstract}

Using the well-known Chern-Weil formula and its generalization, we systematically construct the Chern-Simons forms and their generalization induced by torsion as well as the Nieh-Yan (N-Y) forms. We also give an argument on the vanishing of integration of N-Y form on any compact manifold without boundary. A systematic construction of N-Y forms in D=4n dimension is also given.

\end{abstract}
%\newpage

\baselineskip 18pt
\section{Introduction}
\quad
Curvature plays an essential role in gravitation theory, while torsion is received less attention. However, torsion has also its geometrical meaning and plays some role in gravitation theory.[1-12](for more complete references see [11]) Let us Consider a four dimensional compact nonboundary manifold M with metric $g_{\mu \nu}$. There are two dynamically independent one-forms: the connection $ \omega^a_b$ and vielbein $e^a$ ($g_{\mu \nu}=\eta_{ab}e^a_{\mu}e^b_{\nu}$). We can define curvature and torsion 2-forms out of $\omega^a_b$ and $e^a$ by
\bea
T^a=de^a+\om^a_b \we e^b, \\
R^a_b=d\om^a_b+\om^a_c \we \om^c_b 
\eea
In geometry, $\omm$ and $e^a$ reflect the affine and metric properties of M, while in physics, torsion and curvature may be related to energy momentum tensor and spin current respectively. In spite of the similarity between connection and vielbein, they are different geometrical objects indeed, which can be seen from their local SO(4) transformation properties. Under local transformation $K^a_b$, the connection and vielbein transforms as
\bea
%\begin{array}{l}
 \om \rightarrow K \om K^{-1}+KdK^{-1},\\
e \rightarrow Ke,K \in SO(4),
%\end{array}
\eea
respectively. As we can see from (2), $\omm$ is not a tensor, and $e^a$ is a vector. This difference will have important consequence, which  will be discussed later.

%\quad
Using curvature we can construct characteristic classes, integrations of which will be topological invariants of the manifold, which  reflect the global properties of the manifold. The well known characteristic classes for a 4 dimension real manifold are Pontrjagin and Euler classes [13],
\bea
P=\f {1}{8 \pi^2}  R^{ab} \we R_{ab},\\
E=\f {1}{32 \pi^2}  \va_{abcd} R^{ab} \we R^{cd}
\eea
The integration of $P$ and $E$ on M take integer number, which distinguish topologically different manifolds.

%\quad
It is well known that locally $P_4= dQ$, $Q=\om \we d \om +\f{2}{3} \om \we \om \we \om $ is the local Chern-Simons form, which has many applications in physics. Q is constructed from connection. However we can use the Chern-Weil formula and its generalization to construct global Chern-Simons forms and their generalizations containing torsion. This will be discussed in section 2.
In the same spirit, we construct N-Y form [8-9] in section 3. The N-Y form on 4-d manifold M can be written 
as 
\bea
N=T^a \we  T_a +R_{ab} \we e^a \we e^b ~=dQ_{NY},\\
Q_{NY}=e^a \we T_a
\eea
We will give an arguments on the vanishing of integration of N-Y form on M without boundary. In section 4, we will give a systematic method for constructing N-Y forms in D=4n dimensional manifolds.

\section { Torsion induced Chern-Simons form}
\quad
The main method we will use in this paper is the Chern-Weil homomorphism. Every characteristic class is related to a symmetric, invariant curvature polynomial P and
P satisfies:

(a) dP(R)=0, P(R) is an element of the de Rham cohomology;

(b) the Chern-Weil formula; for any two given connections $\om_0$ and $\om_1$,
 \bea
P(\om_1)-P(\om_0)=dQ(\om_0,\om_1),\\
Q(\om_0,\om_1)=r \int^1_0 P(\eta, \om^{r-1}_t)dt,
\eea
where r is the order of the polynomial P, and
\bea
\om_1 =\om_0+\eta,\\
\om_t=\om_0+t \eta,  0 \leq t \leq 1
\eea
$Q(\om_0,\om_1)$ is the well known Chern-Simons secondary topological class. 

On a 4d manifold $M^4$, there are two typical connections. One is the general affine connection $\om$, the other is the Riemannian connection (torsion free) $\omb$. The curvatures of $\om$ and $\omb$ are R and $\bar{R}$.  The difference of $\om$ and $\omb$ is contorsion.
\bea
\om^a_{b\mu}=\omb^a_{b\mu } +\eta^a_{b\mu },\\
\eta^a_{b\mu }=e^a_{\rho} T^{\rho}_{b \mu}-e^{\rho}_b T^{a}_{\mu \rho}-e^a_{\rho} e^{\la}_b e^c_{\mu} T^{\rho}_{c \la} 
\eea
Applying the Chern-Weil formula (5), we have 
\bea
P(\om)-P(\omb)=dQ_1(\omb,\om), (A_0=\omb, A_1=\om) \\
Q_1(\omb,\om)=tr(2\eta \we \bar{R}+\eta \we \bar{D} \eta +\f {2}{3} \eta \we\eta\we \eta)
\eea    
where $\bar{D}\eta$ is defined by
\be
\bar{D} \eta = d \eta + \omb \we \eta+\eta \we \omb
\ee
It is easy to see that $Q_1$ vanishes whenever $\om$ is torsion free. For a manifold with $\omb=0$, $Q_1=tr(\eta \we d \eta+\f{2}{3} \eta \we \eta\we \eta )$. The form of $Q_1$ is like that of usual local Chern-Simons form, however $Q_1$ is determined completely by torsion and $\int_M dQ_1 =0$. 

Using the generalized Chern-Weil formula,[14-15] we can get the generalized Chern-Simons form. The generalized Chern-Weil formula is,
\bea
(\Delta Q_{k-1})(A_0,...,A_k)=dQ_{k}(A_0,...,A_k)\\
Q_{k}(A_0,...,A_k)=(-1)^{\f{k}{2}(k-1)} \f{n!}{k!(n-k)!} \int_{\Delta^k} P[H^k_0,F^{n-k}(A^0+\sum^{k}_{i=1} t^i(A_i-A_0))]\\
Q_{0}(A_0)=P(F^n(A_0))
\eea
where
\bea
H_0=\sum^k_{i=1} dt^i(A_i-A_0),\\
\Delta^k=\{ (t^1,...,t^k)|\sum^k_{i=1}t^i=1,0 \leq t^i \leq 1 \}\\
\Delta Q_{k-1}(A_0,...,A_k)=\sum^{k+1}_{j=0} (-1)^j Q_{k-1}(A_0,...,\hat{A_j},...,A_k)
\eea
$\Delta$ is also a coboundary operator, $\Delta^2=0$.

For k=0,1, we have tow properties of polynomial P, (a) and (b). For the case k=2, if we choose $ A_0=\omb,  A_1=\om,  A_2=K^{-1}dK$, 
\bea
(\Delta Q_1)(A_0,A_1,A_2)=Q_1(A_0,A_1)-Q_1(A_0,A_2)+Q_1(A_1,A_2)=dQ_2(A_0,A_1,A_2)\\
Q_1(A_0,A_1)=tr[2\eta \we R-\eta \we (d \eta+\om \we \eta +\eta \we \om)+\f{2}{3}\eta \we \eta \we \eta]\\
Q_1(A_0,A_2)=-tr[2\eta_2 \we \bar{R}-\eta_2 \we (d \eta_2+\omb \we \eta_2 +\eta_2 \we \omb)+\f{2}{3}\eta_2 \we \eta_2 \we \eta_2]\\
Q_1(A_1,A_2)=-tr[2\eta_1 \we R-\eta_1 \we (d\eta_1+\om \we \eta_1+\eta_1 \we \om)+\f{2}{3} \eta_1 \we \eta_1 \we \eta_1]\\
Q_2(A_0,A_1,A_2)=2 tr (\eta \we \om -\eta \we K^{-1} dK)
\eea 
where
\bea
\eta_1=A_1-A_2=\om-K^{-1}dK\\
\eta_2=A_0-A_2=\omb-K^{-1}dK
\eea
For the case of k=3,
\be
(\Delta Q_2)(A_0,A_1,A_2,A_3)=dQ_3(A_0,A_1,A_2,A_3)=0
\ee
for arbitrary $A_i$, i=1,2,3,4.

Similar treatment can be applied to Euler characteristic. The only change is the
definition of polynomial in (5).
\bea
E(\om)-E(\omb)=dQ_1(\omb,\om) \\
Q_1 (\omb,\om)= 2 \va_{abcd}(R^{ab} \we \eta^{cd}+\f{1}{2} \bar{D} \eta^{ab} \we
\eta^{cd}+\f{1}{3} (\eta \we {\eta} )^{ab} \we \eta^{cd})
\eea
Again let $ A_0=\omb, A_1=\om, A_2=K^{-1}dK$, we have
\bea
\Delta Q_1(A_0,A_1,A_2)=dQ_2(A_0,A_1,A_2)\\
Q_2(A_0,A_1,A_2)=(2\va_{abcd}\eta^{ab}\we (\om-K^{-1}dK)^{cd})
\eea
and 
\be
(\Delta Q_2)(A_0,A_1,A_2,A_3)=dQ_3(A_0,A_1,A_2,A_3)=0
\ee
for arbitrary $A_i$, i=1,2,3,4.

\section {A simple approach to N-Y form and some comments}
\quad
We now follow the same spirit of section 2 to introduce the N-Y form. We embed the SO(4) connection along with the vielbein into 5 $\times$ 5 skew symmetric matrix [3-7] and define
\bea
W_0=\left(
\begin{array}{cc}
\om&0\\
0&0
\end{array}
\right )\\
W_1=\left(
\begin{array}{cc}
\omm & \f{1}{l} e^a\\
-\f{1}{l}e^b &0
\end{array}
\right )
\eea
where $l$ is a constant with dimension of length.

Though written in the $5 \times 5$ matrix form, $W_0,W_1$ are still SO(4) connections. It can be seen from following. According to (2) W transforms as
\be
W \rightarrow \bar{K}W \bar{K}^{-1}+\bar{K}d\bar{K}^{-1}
\ee 
where
\be
\bar{K}=\left(
\begin{array}{cc}
K^a_b &0\\
0&1
\end{array}
\right )
\ee
The transformation property is just the transformation property of an SO(4) connection. Using Chern-Weil formula again, We have 
\be
P_0(W_1)-P_0(W_0)= l^{-2} dQ_1(W_0,W_1)=\f{N}{l^2}
\ee
where 
\bea
N=T^a \we T_a-R^{ab} \we e_a \we e_b\\
Q=e^a \we T_a
\eea
N is just the N-Y form.

If we let
\bea
W_0=\left(
\begin{array}{cc}
\om&0\\
0&0
\end{array}
\right )\\
W_1=\left(
\begin{array}{cc}
\omm & \f{1}{l} e^a\\
-\f{1}{l}e^b &0
\end{array}
\right )\\
W_2=\left(
\begin{array}{cc}
\omb & 0\\
0 &0
\end{array}
\right )
\eea
where $\omb$ is the Riemannian connection, we have
\be
\Delta Q_1(W_0,W_1, W_2)=Q_1(W_0,W_1)-Q_1(W_0,W_2)+Q_1(W_1,W_2)=dQ_2=0
\ee
where
\bea
Q_1(W_0,W_1)=Q_{NY}=e^a \we T_a\\
Q_1(W_0,W_2)=tr(2 \eta \we \bar{R})+\eta \we \bar{D} \eta +\f{2}{3}\eta \we \eta \we \eta\\
\eta=\om- \omb\\
D \eta=d \eta +\omb \we \eta +\eta \we \omb
\eea
$Q_1(W_0,W_2)$ is just what we have obtained in (8). In fact (24) is still valid if one  substitutes $\omb$ with another arbitrary connection $\om'$ in the definition of $W_2$. 

Several comments are in order.
First, it should be emphasized that $W_1$ is an SO(4) connection rather than an SO(5) connection [10-11]. In fact, if $W_1$ is an SO(5) connection, $e^a_{\mu} e^b_{\nu} \eta_{ab}=g_{\mu\nu}$ is violated and $e^a$ is no long an SO(4) vector. Consequently, the geometric meaning of vielbein and torsion is lost. As a result, we have $\int_M dQ=0$ from Chern-Weil formula. We can arrived the same conclusion in another way. For convenience, we suppose M be $S^4$. We separate $S^4$ into two pieces and identify them with the southern hemisphere $U_S$ and the northern hemisphere 
$U_N$ and $U_S\bigcap U_N=S^3$. We have
\be
\int_{S^4}N = \int_{S^4} dQ 
%=\int_{S^3} Q^+-\int_{S^3}Q_-
\ee
%where
%\be
%Q^+=Q^g_-,~~g \in SO(4)
%\ee
%$Q^g_-$ is the SO(4) gauge transformation of $Q_-$. 
Since $Q=e^a \we T_a $ is a scalar, we obtain the result $\int_{M} N=0$, as the above proof is true in general. 
 So there is no new topological invariant.

Second, N-Y form is a kind of Chern-Simons form and will have its application to manifold with boundary and reflect the role of torsion in geometry. For instance, $Q_{NY}=T^a \we T_a$, when $\om=g^{-1} dg, g\in SO(4) $. In this special case Q is totally determined by torsion.

Third, the generalized Chern-Simons forms constructed in the section 2 are N-Y like forms in the sense that they are all constructed from generalized Chern-Weil formula and induced by torsion and always play essential role on submanifold of corresponding dimension in M.    

\section{Higher dimensional generalization of N-Y form}
\quad
In this section we apply the method further and show how to get the higher dimensional generalization of N-Y form. Though the problem has been considered by O.Chandia and J.Zanelli [10], our treatment is systematic and easier for calculation, furthermore we always make use of the C-W formula which has clearer geometric meaning.

Let us consider the case of D=4n, $n \in Z$, and take D=8 as an example. There are two independent polynomials of order 4. Using them we can construct the following object 
\bea
P=(tr R ^2)^2\\
P'=tr (R^4)
\eea

If we choose two SO(8) connections $\om$ and $\omb$, $\omb$ is the Riemannian connection without torsion. Using Chern-Weil formula, we find $Q(\omb,\om)$ as the Chern-Simons form
\bea
P(\om)-P(\omb)=dQ\\
Q=tr(\eta \we R) \we tr(R \we R)-\f{1}{2}tr(\eta \we D \eta) \we tr (R \we R)-tr( \eta \we R) \we tr(R \we D \eta)\\
+\f{1}{3}[tr(\eta \we \eta \we \eta) \we tr(R \we R)+tr(\eta \we R) \we (2\eta \we \eta \we R)+D \eta \we D \eta)\\
+2tr(\eta \we D \eta) \we tr(R \we D \eta)]+\f{1}{4} [-2tr(\eta \we R) \we tr(\eta \we eta \we D \eta)-\\
2tr(R \we D \eta) \we tr(\eta \we \eta \we \eta)-tr(\eta \we D \eta ) \we tr(2 \eta \we \eta \we R+D\eta \we D \eta)]\\
+\f{1}{5}tr(\eta \we \eta \we \eta)\we tr(\eta \we \eta D\eta) 
\eea
It is straightforward to obtain the generalized Chern-Simons forms.
%Similar result can be obtained for the Euler characteristic.

We may also embed SO(8) connection and frame 1-form $e^a$ into a $9 \times 9$ skew symmetric matrices such as
\bea
W_0=\left(
\begin{array}{cc}
\om&0\\
0&0
\end{array}
\right )\\
W_1=\left(
\begin{array}{cc}
\omm & \f{1}{l} e^a\\
-\f{1}{l}e^b &0
\end{array}
\right )
\eea
where $\om$ is an SO(8) connection. It is easy to check that $W_0,W_1$ are still SO(8) connections. 
Again, using Chern-Weil formula, we have 
\bea
P(W_1)-P(W_0)\\
=4\f{1}{l} tr(R \we R) \we N +4 \f{1}{l^2}N \we N \\
=dQ\\
Q=\f {4}{l}tr(e \we T) \we tr(R \we R)+\f{4}{l^2} tr(e \we T) \we N
\eea
where N is 4d the N-Y form.
Since $l$ is arbitrary constant, we have two exact forms containing torsion
\be
N_1=tr(R \we R) \we  N 
\ee
and
\be
N_2=N \we N
\ee
Same procedure is applied to another polynomial P' as follows:
\bea
P'(W_1)-P'(W_0)\\
=\f{1}{l^2}[4(T_a \we R^{ab} \we e^b)(T_a \we e^a)+(T_a \we T^a)^2-(e_a \we R^{ab} \we e_b)^2]+\\
 \f{1}{l^4}[T_a \we R^{ab} \we R^{bc} \we T_c-e^a \we R^{ab} \we R^{bc} \we R_{cd} \we e^d]\\
=dQ'\\
Q'=\f{4}{l^2}[(e \we T+T \we e)^a_b \we (R \we R)^b_a]+\f{1}{l^4} [ 2tr(e \we T) \we tr(T \we T)+\\
tr(e \we T) \we tr(e \we R \we e)] 
\eea
So, we get the other two closed forms related to torsion,
\be
N_1'=4(T_a \we R^{ab} \we e^b)(T_a \we e^a)+(T_a \we T^a)^2-(e_a \we R^{ab} \we e_b)^2
\ee
and
\be
N_2'=T_a \we R^{ab} \we R^{bc} \we T_c-e^a \we R^{ab}\we R^{bc} \we R_{cd} \we e^d.
\ee
Of course the integrals of above N-Y forms on a manifold without boundary are still zero. But they should play certain role on a manifold with boundary.

Similar result can be obtained for the Euler characteristic.

\section{Conclusion}
\quad
In this paper we use the Chern-Weil formula and its generalization to get some Chern-Simons forms and their generalizations depending on torsion. One of them is the N-Y density and the vanish of integral of N-Y form on a manifold without boundary is proved. We also give a systematic method to construct high dimensional N-Y like forms. The geometric meaning of the generalized Chern-Simons form including N-Y (like) forms on the lower dimensional manifold with boundary and their relation to anomaly are under investigation.\\

{ \large \bf Acknowledgment}\\

This work was supported by Climbing Up Project, NSCC, Natural 
Scientific Foundation of Chinese Academy of Sciences and 
Foundation of NSF in China.
One of the authors, Wei Zhang would like to thank Yi-Hong Gao, Bin Zhou and Yan Luo for helpful discussions.

%\newpage
\begin{flushleft}
\baselineskip 24pt
{\large \bf References}

[1] R.Utiyama, Phys. Rev. 101 (1956) 1597.

[2] J.W.B.Kibble, in Recent Developments in General Relativity, 1962.

[3] H.Y.Kuo (H.Y.Guo), Einstein principle, gauge theory of gravitation and quantum geometrodynamics, Proceedings of the 2nd M.Grossmann Meeting on GR., R.Ruffini(ed.), 475, 1982. 
 
[4] Z.L.Zhou, B.Huang, Y.Z.Zhang, G.D.Li, Y.An, S.Chen, Y.S.Wu, L.L.Zhang, 
Z.X.He and  H.Y.Kuo, Scientia Sinica Vol. XXII No. 6 (1979), 628.

[5] Y.S.Wu, G.D.Li and H.Y.Kuo, Chinese Science Bulletin 19 (1974) 509.

[6] H.Y.Kuo, Chinese Science Bulletin 21 (1976) 31.

[7] Y.An, S.Chen, Z.L.Zhou and H.Y.Kuo, Chinese Science Bulletin, 21 (1976) 379.

[8] H.T.Nieh and M.L.Yan, J. Math. Phys. 23(3) (1982) 373.

[9] H.T.Nieh and M.L.Yan, Ann. Phys. 138(2) (1982) 237.
  
[10] O.Chandia and J.Zanelli, Phys. Rev. D (55) (1997) 7580.

[11] O.Chandia and J.Zanelli, hep-th/9708138 and references therein.

[12] O.Chandia and J.Zanelli, hep-th/9803034.

[13] M.Nakahara, Geometry, Topology and Physics, Adam Hilger, Bristol(1991).

[14] H.Y.Guo, K.Wu and S.K.Wang, Comm. Theor. Phys. (Beijing) 4 (1985) 113.

[15] H.Y.Guo, in Symposium on Anomalies, Geometry, Topology, (1985)  Edited by W.A.Bardeen and A.R.White, World Sci. Pub.

\end{flushleft}
\end{document}